\documentclass[a4paper,fleqn,usenatbib]{mnras}

\usepackage{newtxtext,newtxmath}
\usepackage[T1]{fontenc}
\usepackage{ae,aecompl}

\usepackage{graphicx}	% Including figure files
\usepackage{amsmath}	% Advanced maths commands
\usepackage{amssymb}	% Extra maths symbols

\def\HI{H{\sc i}}
\title[CO in UDGs]{CO observations toward HI-rich Ultra Diffuse Galaxies}

\author[Junzhi Wang et al.]{Junzhi Wang$^{1,2}$\thanks{E-mail: jzwang@shao.ac.cn}, 
Kai Yang$^{2,3}$, Zhi-Yu Zhang$^{4}$, Min Fang$^{5}$, Yong Shi$^{4}$,
\newauthor Shu Liu$^{6}$,  Juan Li $^{1,2}$ and Fei Li$^{1,2}$
\\
$^{1}$Shanghai Astronomical Observatory, Chinese Academy of Sciences,80 Nandan Road, Shanghai, 200030, China\\
$^{2}$Key Laboratory of Radio Astronomy, Chinese Academy of Sciences,   Nanjing, 210008,  China\\
$^{3}$I. Physikalisches Institut, Universit\"at zu K\"oln, Z\"ulpicher Str. 77, 50937 K\"oln, Germany\\
$^{4}$School  of Astronomy and Space Science, Nanjing University, Nanjing,  210093, China\\
$^{5}$Department of Astronomy, California Institute of Technology, Pasadena, CA 91125\\
$^{6}$CAS Key Laboratory of FAST, National Astronomical Observatories, Chinese Academy of Sciences, Beijing 100012, China\\
}

\date{Accepted XXX. Received YYY; in original form ZZZ}

\pubyear{2020}

\begin{document}
\label{firstpage}
\pagerange{\pageref{firstpage}--\pageref{lastpage}}
\maketitle

\begin{abstract}
We present CO observations toward a sample of six \HI-rich Ultra-diffuse
galaxies (UDGs)  as well as one UDG (VLSB-A) in the Virgo Cluster  with the IRAM 30-m telescope.  CO $J$=1-0 is marginally
detected at 4\,$\sigma$ level in AGC122966, as the first detection of CO
emission in UDGs. We estimate upper limits of molecular mass in other 
galaxies from the non-detection of CO lines.  These upper limits and the 
marginal  CO detection in AGC122966 indicate low mass ratios between molecular and 
atomic gas masses. With the star formation efficiency derived from the molecular
gas,  we suggest that  the inefficiency of star formation in such
\HI-rich UDGs is likely caused by the low efficiency in converting molecules from
atomic gas, instead of low efficiency in forming stars from molecular gas.  
\end{abstract}

\begin{keywords}
galaxies: ISM; galaxies: star formation
\end{keywords}

%%%%%%%%%%%%%%%%%%%%%%%%%%%%%%%%%%%%%%%%%%%%%%%%%%

%%%%%%%%%%%%%%%%% BODY OF PAPER %%%%%%%%%%%%%%%%%%

\section{Introduction}

Ultra-diffuse galaxies (UDGs) are extremely low surface brightness galaxies at
the optical and near-IR wavelengths \citep{2017AA...601L..10P}. They have drawn much
attention from both observers and theorists since the
discovery of  47 UDGs in the Coma cluster  \citep{2015ApJ...798L..45V}.
Observations of UDGs  mainly focused on optical and near-IR studies
\citep{2015ApJ...798L..45V,2016AA...590A..20V, 2016ApJS..225...11Y}, which
show that the stellar components are much less than those in normal galaxies.
A sub-set of  UDGs, which have high \HI\, to stellar mass ratio and  
were normally found outside galaxy clusters, have been discovered in \HI\
observations with large single dish radio telescopes, such as Arecibo,
Effelsberg and GBT
\citep{2004MNRAS.352..478R,2017ApJ...842..133L,2018ApJ...855...28S}. 
High angular resolution observations with radio interferometers
\citep{2017ApJ...842..133L,2018AJ....155...65B,2018ApJ...863L...7M,2019AJ....157...76B}
showed that \HI\, emission are more extended than the stellar components traced
by optical images. The line widths (FWHM) of \HI\ detected by single dishes
\citep{2004MNRAS.352..478R,2017ApJ...842..133L,2018ApJ...855...28S}  and
interferometers
\citep{2017ApJ...842..133L,2018AJ....155...65B,2018ApJ...863L...7M}  in such
galaxies are only about or even less than 100 km s$^{-1}$. Such small
linewidths led difficulties for determining dynamical mass of whole galaxy with
large uncertainty of inclination angle. 

With extremely high ratios of gas to stellar mass and with low stellar mass
\citep{2017ApJ...842..133L, 2018ApJ...863L...7M}, such \HI-rich UDGs  should
have inefficient star formation, low star formation rate (SFR), and low star
formation efficiency (SFE) across cosmic time. \HI-rich  UDGs with a size of
the Milky Way may be ``failed" $L_{\star}$ galaxies or ``failed" smaller
galaxies, depending on the estimated halo mass  \citep{2017ApJ...842..133L}.

There are two major steps to form stars from \HI~  gas: atomic gas converts 
to molecular gas, and molecular gas form stars. So, molecular to atomic gas
fraction is a key parameter to understand the inefficient star formation in
such galaxies. However, the lack of molecular gas information of such galaxies prohibits further distinction about 
which step is more crucial for the 
inefficient star formation in the history. Therefore, CO observations toward a sample of \HI-rich
local UDGs would help answer such questions:  
The inefficient star formation in such UDGs is caused by   difficulties of forming
molecular gas from atomic gas, or failed star formation from molecular gas?

Although there have been several   CO observations with few detections   \citep{2000ApJ...545L..99O, 2001ApJ...549L.191M, 2005AJ....129.1849M, 2010A&A...523A..63D,   2017AJ....154..116C} 
  toward \HI\, rich Low Surface Brightness Galaxies (LSBGs), which have quite high gas to star mass ratios, there is still no report on CO observations toward UDGs in the literature, up to now.

In this letter, we describe CO line observations and data reduction of a small
sample of UDGs with the IRAM 30-m telescope in \S2, present the main results
and discussions in \S3, and make the brief summary and future prospects in \S4.

\section{Observations and data reduction}

 We select six \HI-rich UDGs from the literature \citep{2017ApJ...842..133L,2017AA...601L..10P,2017MNRAS.467.3751B,2017ApJ...836..191T,2019AJ....157...76B} to perform CO line
observations,
because they have enough gas materials to form stars.  The sample includes
three Milky-Way-sized ones with \HI\, mass $\ge$10$^9M_\odot$, three dwarf
galaxies with \HI\, mass $<$ 2.5$\times$10$^8M_\odot$ (see Table 1), and
VLSB-A, which is one of the three UDGs in the Virgo Cluster with velocity
measured through optical spectra toward the nucleus
\citep{2015ApJ...809L..21M}.

The
observations were carried out from July 31 to August 3, 2018  with  the IRAM 30-m
millimeter telescope at Pico Veleta, Spain\footnote{Based on observations
carried out with the IRAM 30m Telescope.  IRAM is supported by INSU/CNRS
(France), MPG (Germany) and IGN (Spain).}.  The Eight MIxer Receiver (EMIR)
with dual polarizations, FTS backend, and  standard wobbler switching mode with
$\pm120''$ offset at 0.5 Hz beam throw, were used. Focus calibrations were done at the
beginning of the observations and during sunset or sunrise, toward planets or
strong millimeter quasars. Pointing calibrations were done every 2 hours using nearby
quasars.  CO 1-0 at E0 band and CO 2-1 at E2 band were covered simultaneously
our observations. The observing frequency of CO 1-0 varies from $\sim$112.35 GHz with typical system temperature of 150K  in SECCO-dI-1 to $\sim$115.273 GHz with  typical system temperature of 300K  in VLSB-A.  The  system temperature also varies with different weather conditions and elevations of telescope. 
One the other hand, CO 2-1 observations are from $\sim$ 224.7 GHz to 230.5 GHz, while the typical system temperature is about 500K and strongly depends on   weather conditions and elevations of telescope.

  The pointing center of each source is listed in Table
1. The optical images of these sources  from the PTF (the Palomar
Transient Factory) \citep{2009PASP..121.1395L, 2009PASP..121.1334R}  or
the ZTF (the Zwicky Transient Facility) \citep{2019PASP..131a8002B} are
presented in Figure 1 and 2.

The data were dumped every 1.7 minutes as one scan, and calibration was done
every 6 scans. The beam sizes of the IRAM 30-m telescope are about 22$''$ for
CO 1-0 and 11$''$ for CO 2-1, respectively.  The conversion from  $T_{\rm
A}^*$  to $T_{\rm mb}$ is: $T_{\rm mb}=T_{\rm A}^*F_{\rm eff}/B_{\rm eff}$,
where $F_{\rm eff}$=95\% and $B_{\rm eff}=81\%$  for CO 1-0, while  $F_{\rm
eff}$=92\% and $B_{\rm eff}=59\%$  for CO 2-1.

\begin{figure}
%	\includegraphics[width=\columnwidth]{example}
\includegraphics[width=0.45\textwidth]{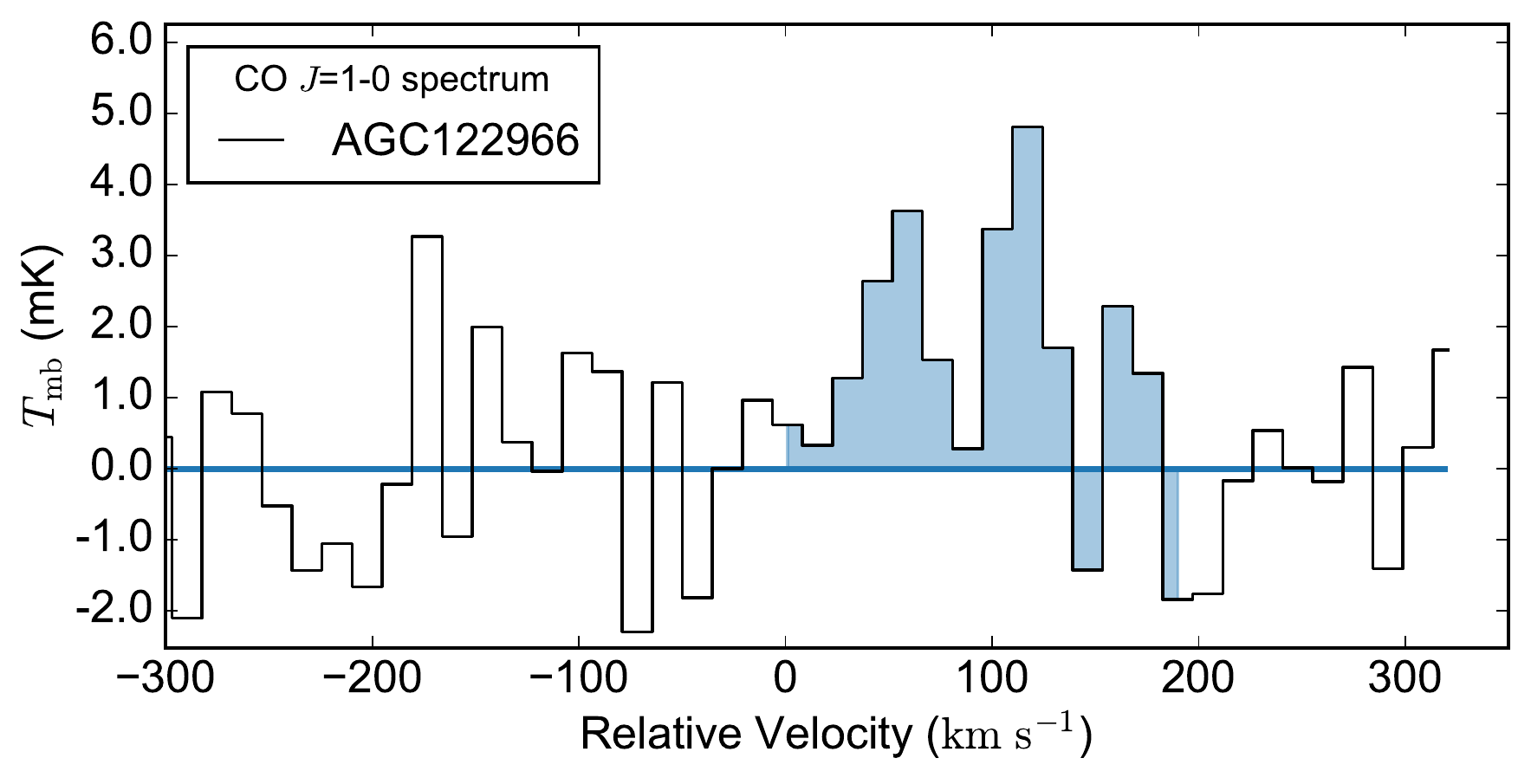}
\includegraphics[width=0.45\textwidth]{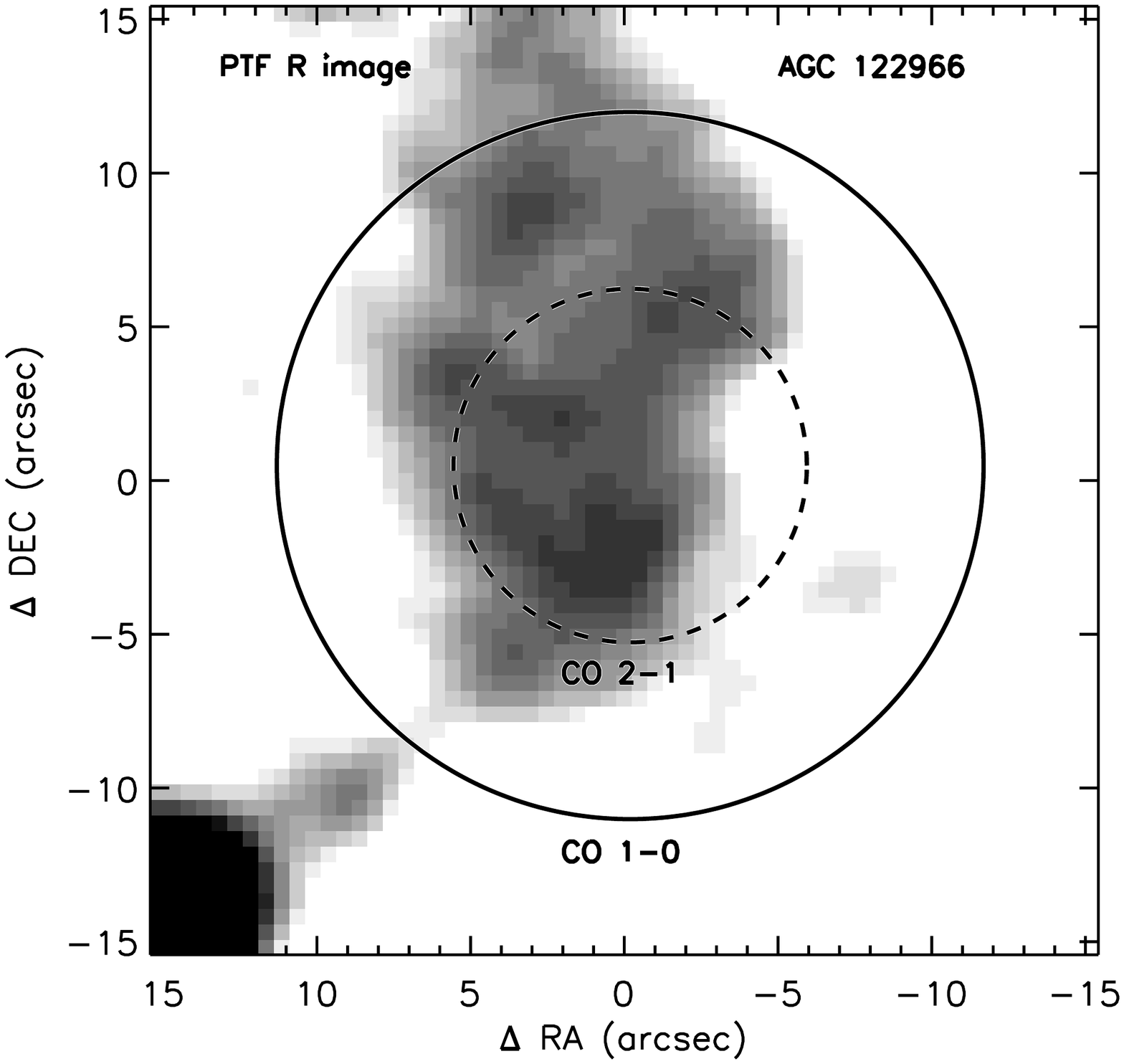}

\caption{Top: CO 1-0 spectrum of AGC122966 obtained with IRAM 30 meter, with
        velocity resolution of 14.53 km s$^{-1}$. Bottum: The beams of IRAM 30
        meter telescope for CO 1-0 (solid circle) and 2-1 (dashed circle)
        overlaid on the PTF (the Palomar Transient Factory)
        \citep{2009PASP..121.1395L, 2009PASP..121.1334R} R band of AGC122966.}

    \label{fig:figure1}
\end{figure}

The CLASS package of GILDAS\footnote{http://www.iram.fr/IRAMFR/GILDAS} was used
for data reduction.  The effective on-source time and the noise level for each
source are listed in Table 2. First order baseline fitting was done for each
spectrum with 1.7 minutes integration time.  Then we smoothed and resampled
each spectrum to $\sim$ 10 $\rm km\,s^{-1}$, before averaging all spectra of
each target (Fig. 1) with a weighting of $1/T_{\rm sys}^2$.  For each target,
we calculated the standard deviation of each channel across all spectra.  This
gives a channel based noise, which is pretty flat for all sources and 
is consistent with  the $rms $ value  from the final spectrum  obtained with standard method using   class
task `average' f in each source. Even though the rms noise increases with the increasing of the frequency at 3mm band due to the O$_2$ line in the earth's atmosphere, it is almost the same within $\pm$500 km s$^{-1}$ for the observed CO 1-0 line.

 The noise level ($\sigma$) in $T_{\rm mb}$ for each source
is listed in Table 2 are obtained from the averaged spectrum at the original frequency
resolution (0.195 MHz),  with line free channels using CLASS task `base' with first order polynomial. The upper limits of flux are calculated with
3$\sigma\times\sqrt{{\delta}v\times{\Delta}V}$, where $\sigma$ is the rms noise from
the final spectrum for each source with 0.195 MHz resolution,  ${\delta}v$ is
the velocity resolution corresponding to 0.195 MHz ($\sim0.5$ km s$^{-1}$), and
${\Delta}V$ is 50 km s$^{-1}$, as the assumed line width.  For most of the sources,
since the observations were done in summer time, CO 1-0 data were much better
than CO 2-1 due to weather conditions. Thus, CO 1-0, instead of CO 2-1 data,
are used for estimation of the upper limits.

\begin{figure}
\includegraphics[width=0.25\textwidth]{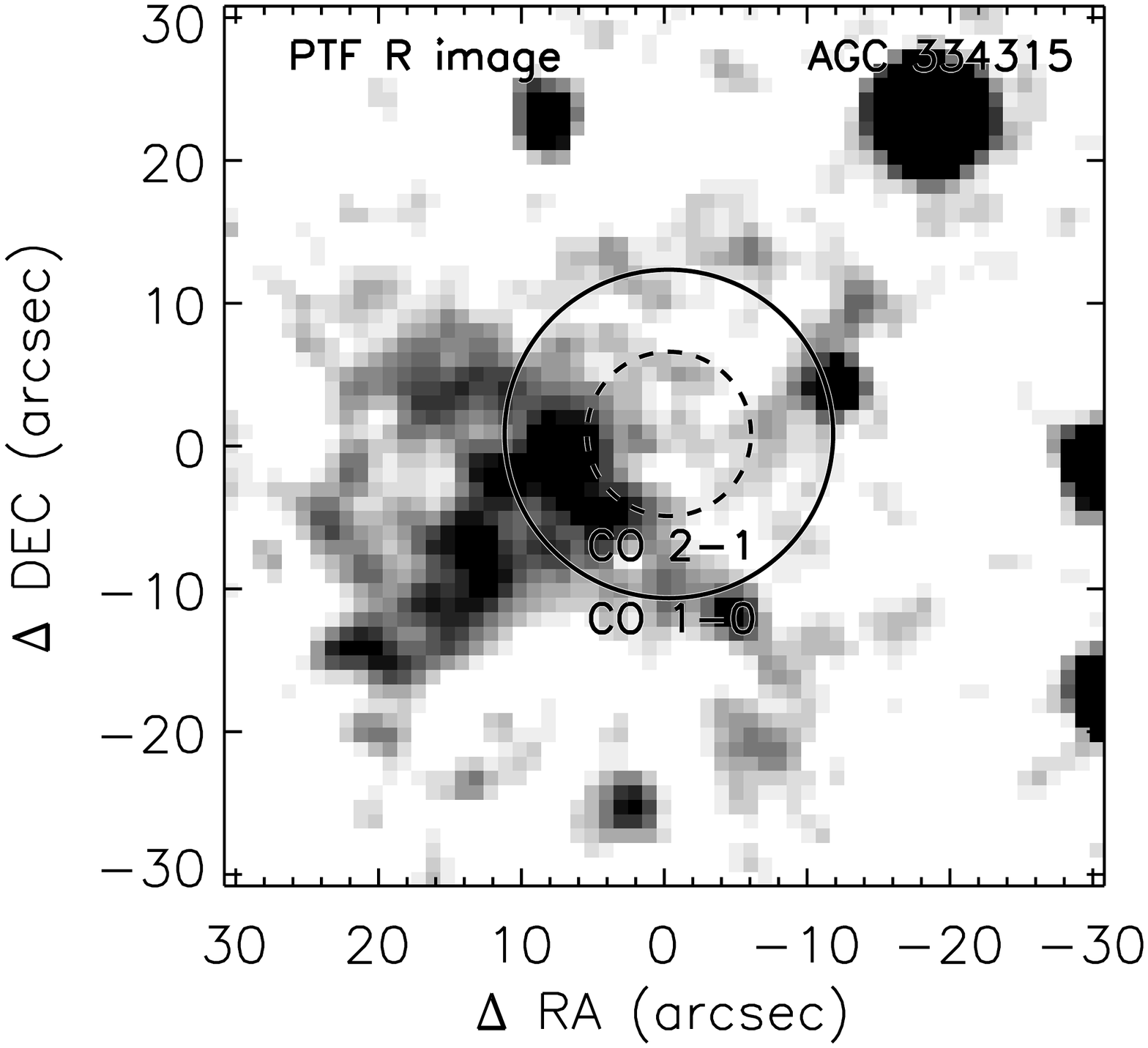}%SECCO-dI-2.eps}
\includegraphics[width=0.25\textwidth]{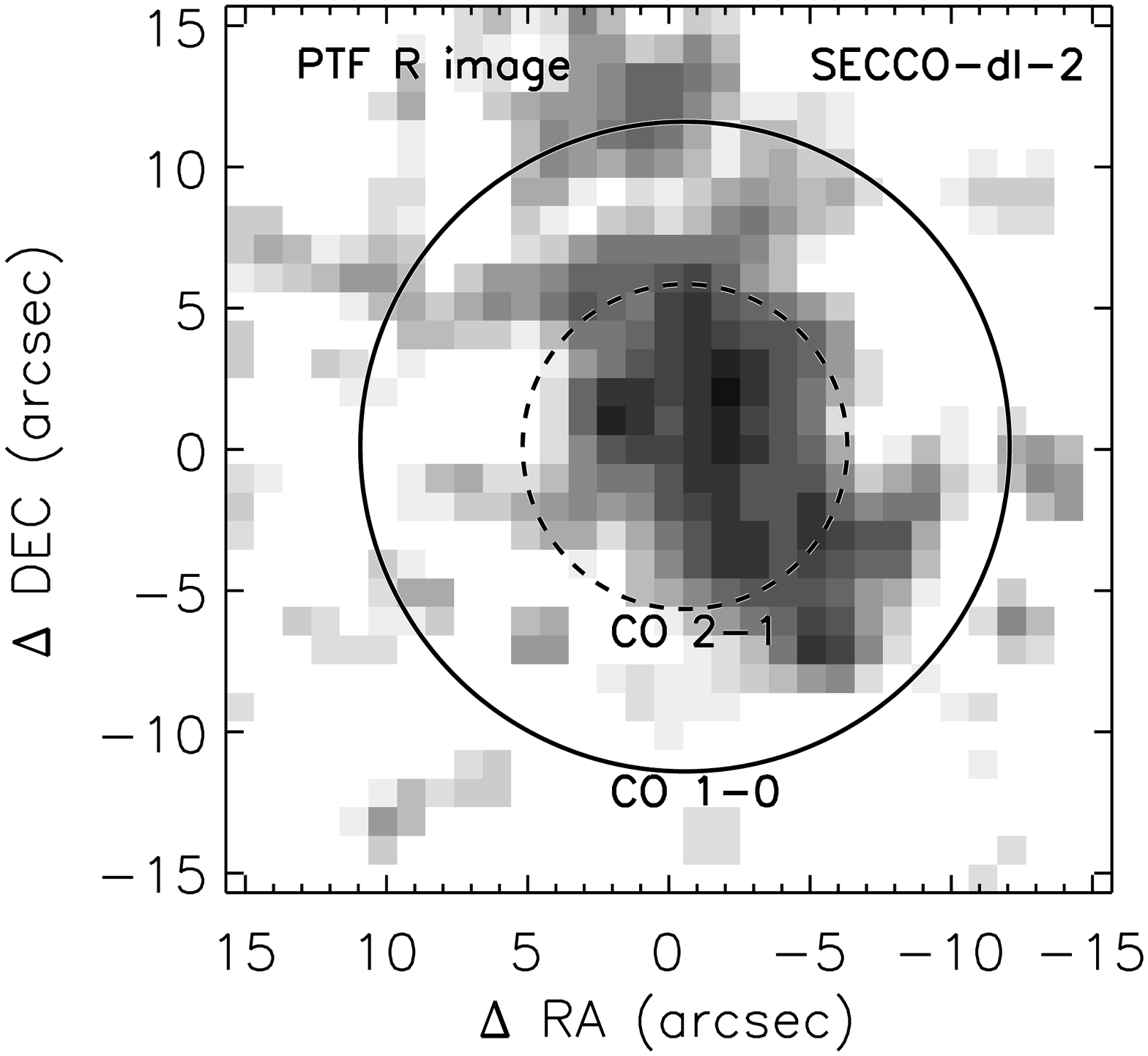}
\includegraphics[width=0.25\textwidth]{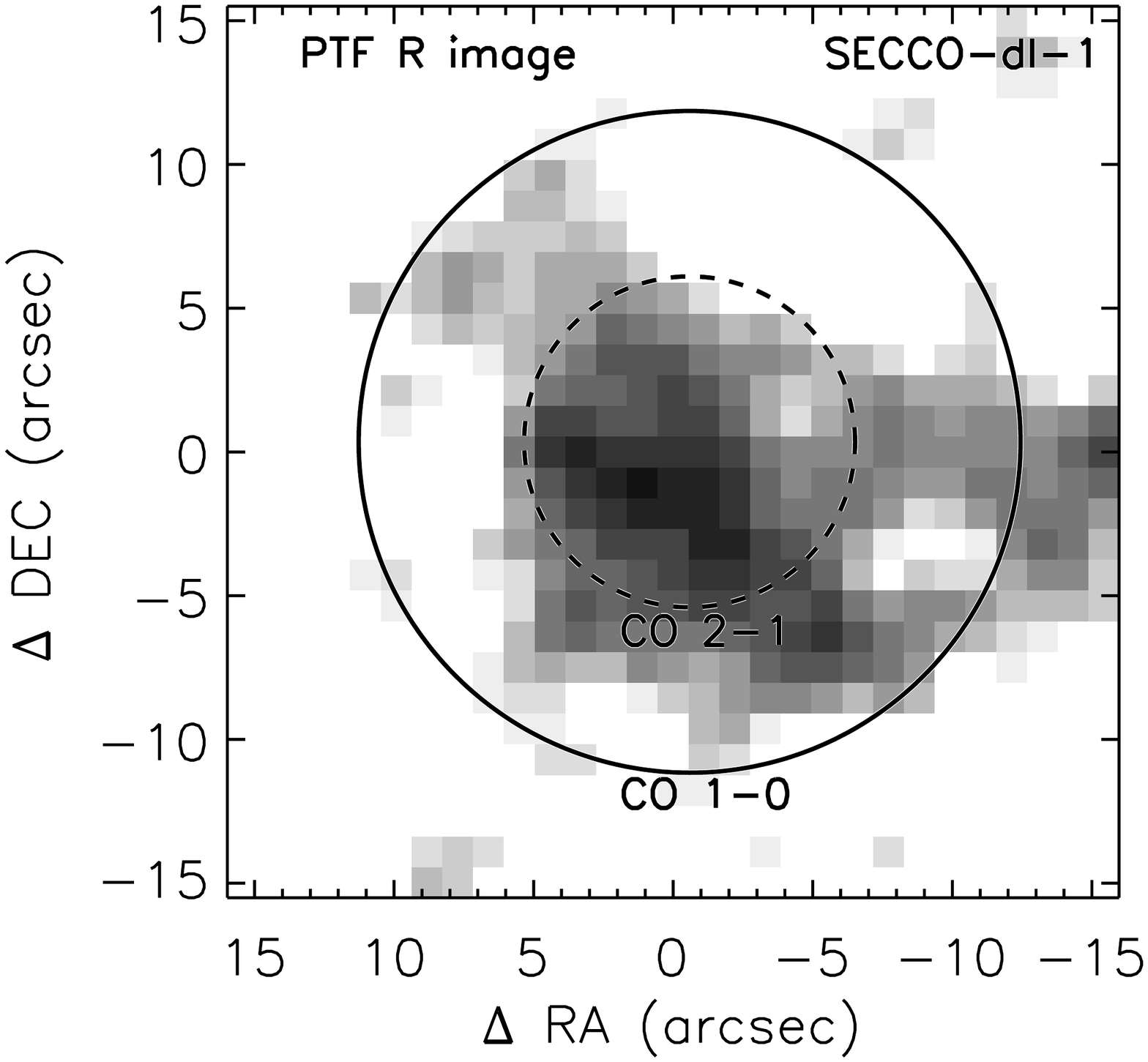}\includegraphics[width=0.25\textwidth]{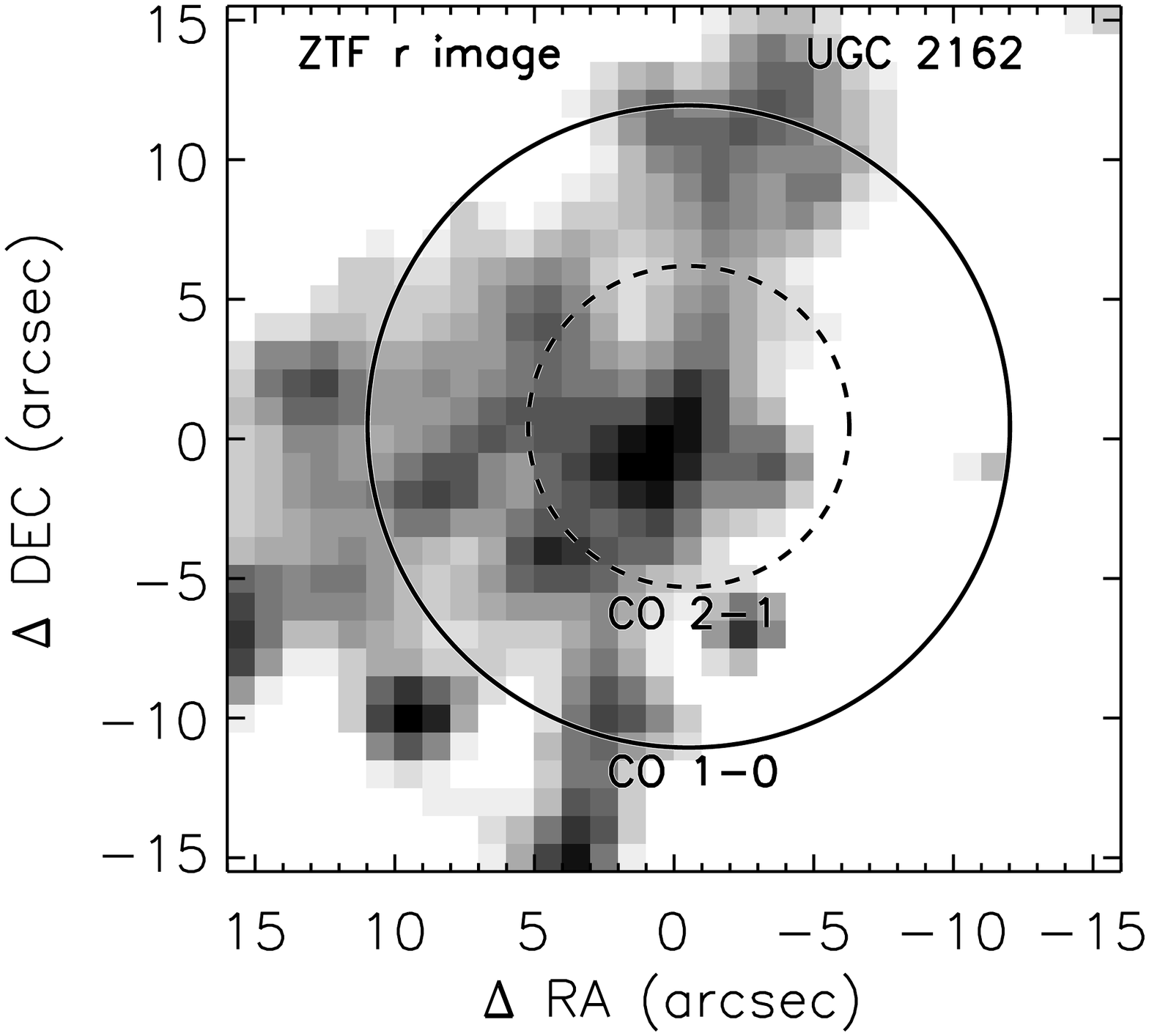}
\includegraphics[width=0.25\textwidth]{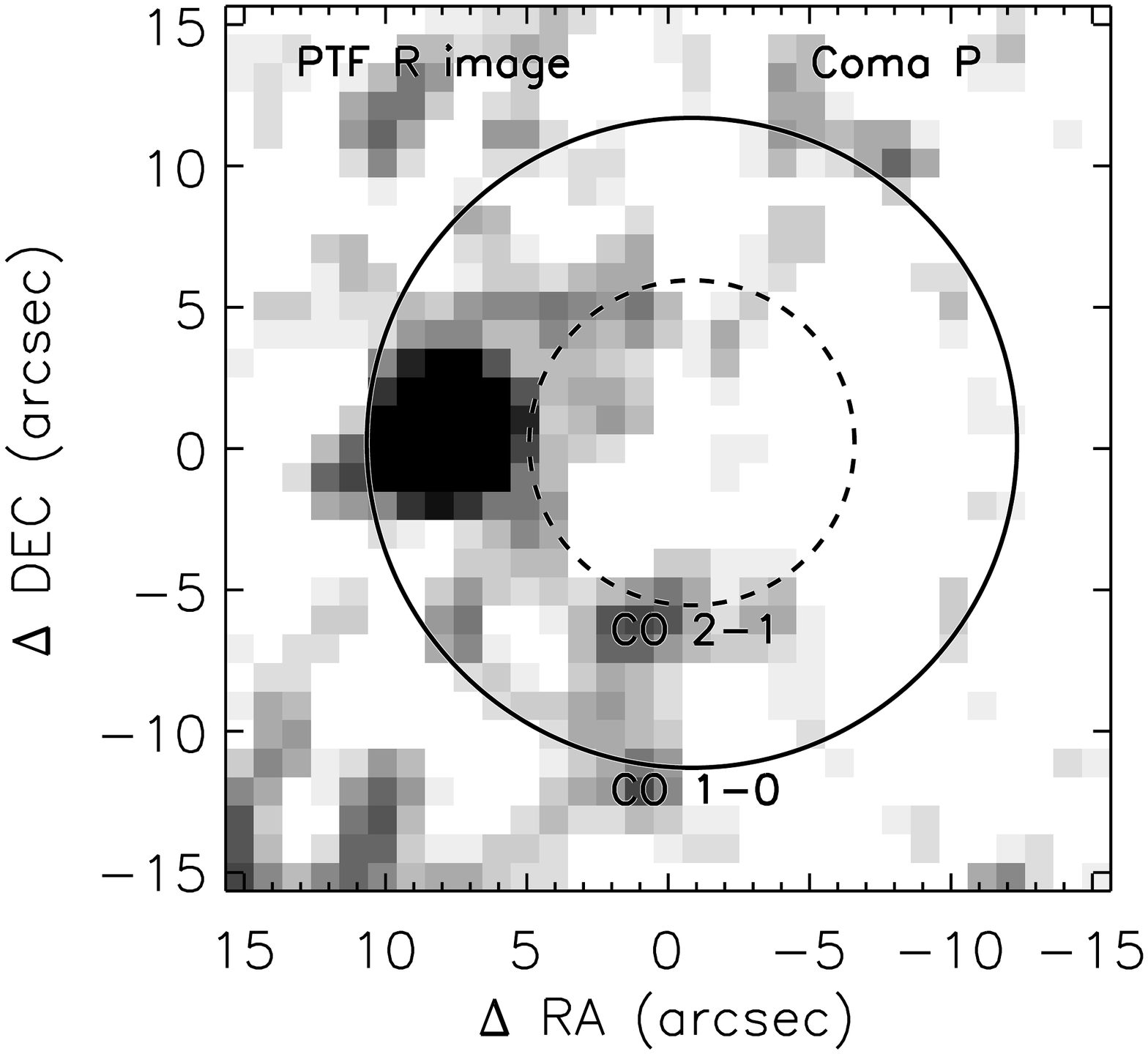}\includegraphics[width=0.25\textwidth]{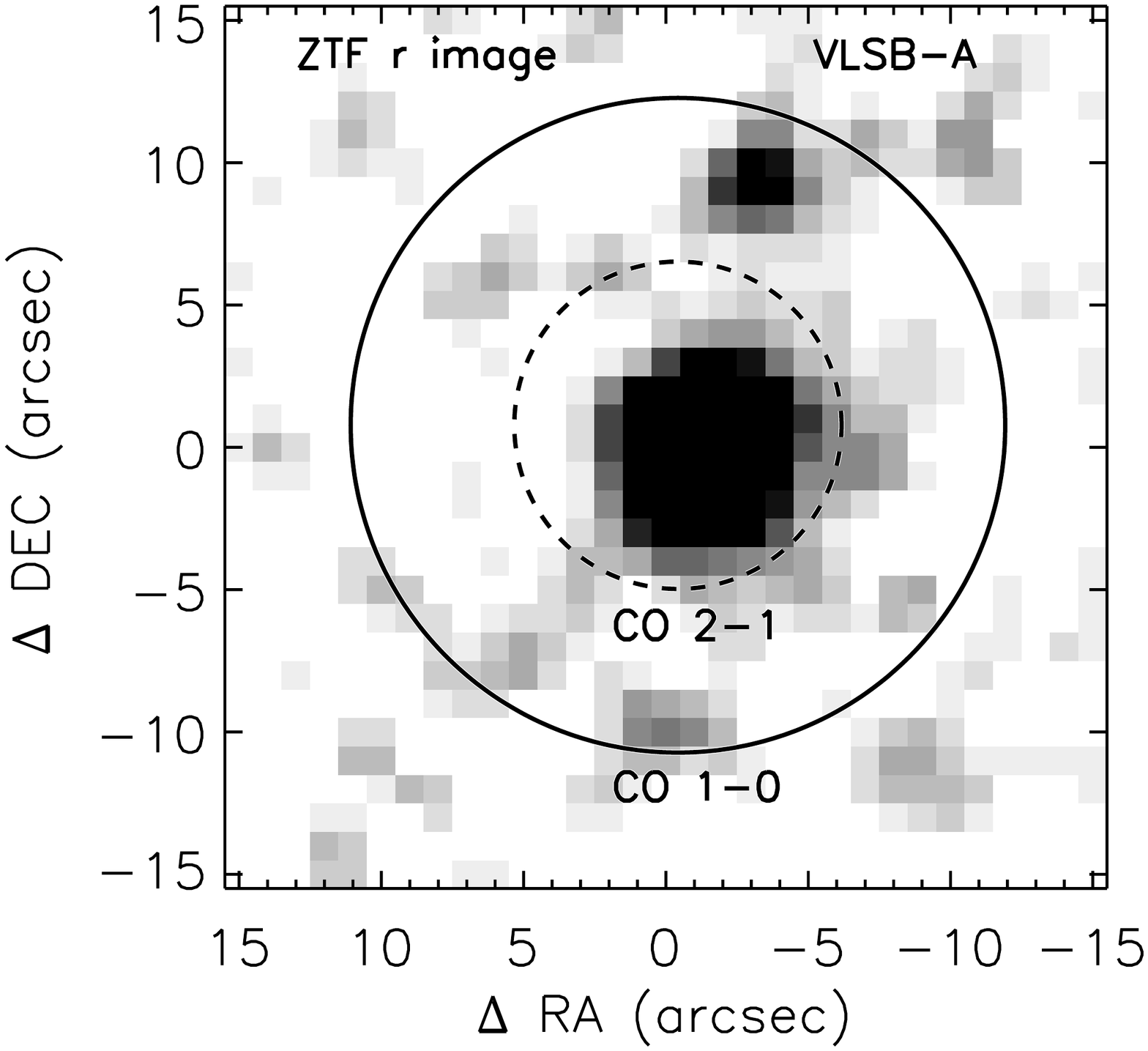}
    \label{fig:figure2}

\caption{The beams of IRAM 30 meter telescope for CO 1-0 (solid circle) and 2-1
        (dashed circle)  overlaid on the PTF (the Palomar Transient Factory)
        \citep{2009PASP..121.1395L, 2009PASP..121.1334R} R band  or ZTF (the
        Zwicky Transient Facility) \citep{2019PASP..131a8002B} r band images of
individual UDGs. }
\end{figure}

\section{Results and Discussion}

CO 1-0 emission is marginally detected in AGC122966 at 4 $\sigma$ level
with velocity integrated flux of 0.23$\pm0.057$ K km s$^{-1}$ in $T_{mb}$ (see
Figure 1), while only upper limits at can be estimated in  other sources. The
velocity range of CO 1-0 emission in AGC122966 is from about $+$50 to $+$150
km s$^{-1}$, relative to the systematic velocity obtained from the \HI\ 21-cm
emission.

The CO 2-1 line of AGC122966, which was obtained simultaneously, was not
detected at 3$\sigma$.  The non-detection of CO 2-1 may caused by the high
system temperature at the 1.3 mm band. Further more, CO 2-1 has smaller beam
size, which is only 25\% of the CO 1-0 beam coverage. So, if the CO emission is
clumpy and located off the pointing center of this observation, it is possible
that CO clump is not fully covered by the CO 2-1 beam.  To present the
difference of the beam sizes, we overlay the beams on the stacked R-band images
of AGC122966, which are from the intermediate Palomar Transient Factor (PTF)
survey, and show it in Figure 1.

The  velocity  integrated luminosity of CO 1-0 ($L_{CO}$) in AGC122966 is
2.2$\pm0.54$ K km s$^{-1}$ pc$^{-2}$,  using the formula
$\frac{\pi}{4\rm{ln}2}\theta_{mb}^2 I \times D_L^2\times(1+z)^{-3}$
\citep{1997ApJ...478..144S,2004ApJS..152...63G,2015ApJ...799...92J},  where
$\theta_{mb}$=22$''$ is the main beam size of the observation, $I$ is the
velocity integrated flux in  $T_{\rm mb}$, $D_{\rm L}$ is the luminosity
distance of the galaxy, and $z$ is the redshift. The  CO 1-0 emission is
out of \HI\, velocity range \citep{2017ApJ...842..133L,2019ApJ...883L..33M}
within $\pm$45 km s$^{-1}$. The marginal detection of CO 1-0 might not be real
 because of the different velocity ranges of \HI\, and CO 1-0.  However, even though no related  \HI\ 21-cm line emission at the same velocity range  as that of the  CO 1-0 peak are reported, weak \HI\, 21-cm below the detection limit in the literature are still possible, which means that the CO 1-0 detection
can still be real. Last, even if the 3$\sigma$ upper limit of CO 1-0 is
adopted to estimate molecular gas mass of AGC122966, instead of using the
detection, the main result is still  similar to considering this feature
as detection.

The metallicity in UDGs is difficult to be measured due to weak optical
emission. Only SECCO-dI-2  and UGC 2162 have such measurements in the
literature, which gave low metallicities of 8.2$\pm0.2$ of  12+log(O/H) for
SECCO-dI-2 \citep{2017MNRAS.467.3751B} and 8.22$\pm0.07$ of 12+log(O/H)  for
UGC 2162 \citep{2017ApJ...836..191T}, respectively.   SECCO-dI-2 and
UGC 2162 do not follow the mass-metallicity relation in galaxies
\citep{2012ApJ...754...98B}, which gave the metallicity of $\sim$7.6 for
12+log(O/H) for galaxies with similar mass to SECCO-dI-2 as
$M_\ast$=$0.9\times10^7M_\odot$ \citep{2017MNRAS.467.3751B} or  UGC 2162 as
$M_\ast$=$2.0\times10^7M_\odot$  \citep{2017ApJ...836..191T}. A tentative
evidence that the gas-phase metallicities in  diffuse systems are high for
their stellar mass had been found in two UDGs \citep{2018ApJ...866..112G}.
With limited information of metallicity  measurement toward these UDGs, we
would like to assume similar metallicity to  that of SECCO-dI-2  in these UDGs
in our discussion.

For galaxies with a metallicity of 8.2$\pm0.2$,  the conversion factor
($\alpha_{CO}$), which converts from CO luminosity to molecular mass, should be
higher than that in the Milky Way, base on the relation of metallicity and
$\alpha_{CO}$ \citep{2013ARA&A..51..207B, 2016NatCo...713789S}.  $\alpha_{CO}$
(8.6$M_{\odot}$/K km s$^{-1}$pc$^{2}$ ),  with twice of that in the Milly Way
 as a reasonable value for such metallicity, is adopted (see Table 2).
The molecular gas mass in AGC122966 is then estimated to be
1.9$\pm0.46\times10^8M_{\odot}$,  if the emission feature is real. The
H$_2$/\HI\,  mass ratio is 0.19$\pm0.05$, while it ranges from $\sim$0.03 to 3
in a large sample of galaxies \citep{2015ApJ...799...92J}.

Using the same method as for AGC122966, the non-detection of CO emission in
most of these \HI-rich UDGs can give 3$\sigma$ upper limits of
molecular gas mass, assuming a line width of 50 km s$^{-1}$ in each
galaxy.  Upper limits of H$_2$/\HI\  mass ratios for each galaxy can also be
estimated (see Table 2), which are from 0.012  in UGC 2162 to 0.083 in
SECCO-dI-1.  The upper limit of molecular gas mass is also estimated for AGC
122966 if that emission feature is not real. The 3 $\sigma$ upper limit of CO
1-0 in AGC 122966 is about 1/4 of the flux of that feature, not only because
that feature is only 4$\sigma$, but also because the line width of that feature
is more than 50 km s$^{-1}$.  
Note that the pointing centers of our observations are  with about 15$''$ and 9$''$  offsets    from the optical centers in AGC 334315 and ComaP (see Figure 2.), respectively.
 The offset for AGC 334315 is about  70\%  beamsize of  CO 1-0, while it is about 40\% for  ComaP.  Most of the emission at optical band in ComaP can be covered with the beam of CO 1-0 observation. However, only 1/3 to 1/2 of the optical emission region in AGC 334315   can be covered by CO 1-0 beam. Such offset can cause underestimation of CO  upper limit in these two galaxies,  which  can further cause about a factor of 2 in ComaP    or  3 in AGC 334315  underestimation of H$_2$/HI ratio and overestimation of SFR/$M_{H_2}$.

 The ratio of molecular mass and stellar mass, $M_{H_2}/M_\ast$, in each
galaxy, is also derived and listed in Table 2. The upper limits of this ratio
range from 0.12 to 10, while this ratio ranges from $\sim0.01$ in massive
galaxies with $M_\ast\sim10^{11.5}M_\odot$ to  $\sim1$ in low mass end with
$M_\ast\sim10^{8.5}M_\odot$, decreasing with  M$_\ast$ at the massive end 
and flatten at the low mass end with large scatter
\citep{2015ApJ...799...92J}. No clear difference of $M_{H_2}/M_\ast$ between
\HI-rich UDGs to normal galaxies  can be justified with current results. On the
other hand,  the ratios of $M_{ HI}/M_{\ast}$ in these HI-rich UDGs, range
from 8.3 to 120, are much higher than that in normal galaxies ranging from
$\sim0.01$ at massive part to almost ten in the low mass end
\citep{2015ApJ...799...92J}.

The low H$_2$/\HI\, mass ratios, i.e., the lack of molecular gas,  in these
\HI-rich UDGs,  all of which are less than 0.1 (see Table 2), might be the main
reason of inefficient star formation,  while such ratio in normal galaxies are from $\sim0.03$ to $\sim3$ with a median value of $\sim0.3$ \citep{2015ApJ...799...92J}.  If only the molecular gas
was used for calculating  SFE   (SFR/$M_{H_2}$), such value will not be
significantly lower than that in nearby spirals as
(5.25$\pm2.5)\times10^{-10}$yr$^{-1}$  \citep{2008AJ....136.2782L}. The SFR
in AGC122966, the one with tentative CO 1-0 detection,  is 0.022 $M_{\odot}$ yr$^{-1}$
\citep{2017ApJ...842..133L},  which gives an SFE of 
1.3$\times10^{-10}$yr$^{-1}$.  On the other hand, with an SFR of 0.045
$M_{\odot}$ yr$^{-1}$ \citep{2017ApJ...842..133L} and a non-detection of CO 1-0
in AGC 334315, the derived SFE with molecular gas is greater than 12.1
$\times10^{-10}$yr$^{-1}$, which is much higher than that in nearby spirals.
Therefore, we suggest that the inefficient star formation in such galaxies
is mainly due to the low efficiency of forming molecules from atomic gas, instead
of forming stars from molecular clouds. 

 With limited information of metallicity in these UDGs,  the adopted
$\alpha_{CO}$ may be lower than the real value, which would underestimate
molecular gas mass. In this case, low SFE for molecular gas can not be the only
explanation of inefficient star formation in such \HI-rich UDGs.
Deep CO observations toward UDGs with ALMA,  as well as  metallicity
measurement with deep optical spectroscopic observations, will help us to
derive molecular gas mass and determine the reason of inefficient star
formation.

Since molecular gas and stellar masses are much less than the atomic mass in
such galaxies, the baryonic mass can be estimated with only \HI, to derive the
dynamical-to-baryonic mass ratios.  As discussed in the literature
\citep{2017ApJ...842..133L,2019AJ....157...76B}, the total masses estimated
with dynamics are not massive enough, such \HI-rich UDGs should not be failed
$L_\star$  spiral galaxies, even though some of them have \HI\ and optical
sizes similar to those of $L_\star$  spiral galaxies. 

The properties of such galaxies, such as high halo spin parameters
\citep{2017ApJ...842..133L}, may bring down the efficiency of forming molecules
from atomic gas.  AGC122966, with marginal detection of CO 1-0 emission, is the
galaxy with lowest $M_{HI}$/$M_{*}$ ratio (see Table 1) among the 6 \HI-rich
UDGs.  The relatively high H$_2$/\HI\ ratio in AGC122966 may cause higher star
formation rate than other \HI-rich UDGs in the past and nowadays, which can
further explain the relatively low $M_{HI}$/$M_{*}$ ratio in AGC122966. 
Or, if that emission feature is not real, AGC122966 will be similar to
other UDGs. Further confirmation of CO emission in AGC122966 with millimeter
interferometers, such as ALMA or NOEMA, is necessary to make a conclusion.

\section{Summary and future prospects}

As the first survey observations of CO 1-0 and 2-1 toward a sample of \HI-rich
UDGs with the IRAM 30-m telescope, we  marginally detected CO 1-0 in one
galaxy (AGC122966) at 4$\sigma$ level. The non-detection of CO lines in other
sources provides good upper limits of molecular masses, which help 
estimate upper limits of molecular gas mass. We find low ratios of molecular to
atomic mass, which indicate that the inefficient star formation in such \HI-rich
UDGs should mainly be caused by the difficulty of forming molecules from atomic
gas, while the star formation efficiency derived for molecular gas is not
significantly lower than normal spirals. An alternative possibility is
that CO lines are no longer good tracers of molecular gas in UDGs,  because the metallicity might be lower than assumed.

Further large-sample high-sensitivity CO observations with ALMA can better
derive molecular mass in such galaxies and can better provide the molecular to
atomic mass ratios.  ALMA observations can also provide the spatial
distribution of molecular gas, which can be used to compare with star formation
informations using further optical emission line observations.


\section*{Acknowledgements}

We thank the referee, Dr.  U. Lisenfeld,  for helpful suggestions to improve the manuscript.  
This work is supported by  National Key Basic Research and Development Program
of China (grant No. 2017YFA0402704) and the National Natural Science Foundation
of China grant 11590783, and U1731237.  This study is based on observations
carried out under project number 068-15 with the IRAM 30-m telescope. IRAM is
supported by INSU/CNRS (France), MPG (Germany) and IGN (Spain). This work also
benefited from the International Space Science Institute (ISSI/ISSI-BJ) in Bern
and Beijing, thanks to the funding of the team ``Chemical abundances in the
ISM: the litmus test of stellar IMF variations in galaxies across cosmic time"
(Principal Investigator D.R. and Z-Y.Z.). J.W. thanks Dr. Yu Lu for helpful
discussion about UDGs.

\section*{Data availability}
The original CO  data observed with IRAM 30 meter  can be accessed by IRAM archive system at https://www.iram-institute.org/EN/content-page-386-7-386-0-0-0.html, while optical images can be obtained from PTF and  ZTF archive system. If anyone is interested in the  reduced  data presented in this paper, please contact  Junzhi Wang at jzwang@shao.ac.cn.

%%%%%%%%%%%%%%%%%%%%%%%%%%%%%%%%%%%%%%%%%%%%%%%%%%



%\bibliographystyle{mnras}


%%%%%%%%%%%%%%%%%%%%%%%%%%%%%%%%%%%%%%%%%%%%%%%%%%

%%%%%%%%%%%%%%%%% APPENDICES %%%%%%%%%%%%%%%%%%%%%

%\appendix

\clearpage
\begin{table}
%	\centering
\begin{center}
	\caption{Source list}
	\label{tab:table1}
	\begin{tabular}{cccccccccc} % four columns, alignment for each
		\hline
		Source name & RA(J2000) & DEC(J2000) & $cz$ &Dist&$r_e$&$M_{HI}$&$M_{HI}/M_{\ast}$&SFR&References\\
		                           &                      &                   &  km s$^{-1}$&Mpc&kpc&10$^9M_{\odot}$&&$M_{\odot}yr^{-1}$\\
		\hline
AGC 122966 &02:09:29.0 &+31:51:15.0&6518&90&7.4$\pm3.3$&1.0&8.3&0.022&1\\
%AGC 219533  & 11:39:57.0  & +12:43:10.0      & 6380 &96 & 3.7$\pm0.7$&1.6&24&0.030&1\\
SECCO-dI-1&11:55:58.5&+00:02:36.3 &7791&112& 2.6 &1.2 &120 & ... & 2\\
AGC 334315&23:20:11.0 &+22:24:10.0&5100&73 &4.2$\pm1.1$&1.4&23&0.045 &1\\
\hline
UGC 2162&02:40:23.1&+01:13:45.0&1172 &12.3 &1.7 &0.19&10&8.7$\times10^{-3}$ & 3 \\
SECCO-dI-2 &11:44:33.8&-00:52:00.9&2543 &40 &1.3& 0.24&27  &... &4\\
Coma P&12:32:10.3 & +20:25:23.0&1348& 5.5&$<$1&0.035&81& 3.1$\times10^{-4}$& 5 \\
\hline
VLSB-A & 12:28:15.9&+12:52:13.0&-120& 16.5&9.7&...&...&... & 6\\
		\hline
	\end{tabular}
	\end{center}
The sources are separated into 3 sub-groups: Milky Way size HI-rich UDGs (AGC 122966, SECCO-dI-1 and AGC 334315), dwarf  HI-rich UDGs (UGC 2162, SECCO-dI-2 and  Coma P), and one UDG without HI detection in the Virgo Cluster. \\
References:1. \cite{2017ApJ...842..133L}, 2.  \cite{2017MNRAS.467.3751B}, 3. \cite{2017ApJ...836..191T}, 4.  \cite{2017AA...601L..10P}, 5.  \cite{2019AJ....157...76B}, 6.  \cite{2015ApJ...809L..21M}
\end{table}

\begin{table}
%	\centering
\begin{center}
	\caption{Observational results}
	\label{tab:table1}
	\begin{tabular}{cccccccccc} % four columns, alignment for each
		\hline
		Source name &on-source time& rms$^a$ & $I$(CO 1-0)$^b$ &$L$(CO) &$M_{H_2}$$^d$  &$M_{H_2}/M_{HI}$&$M_{H_2}/M_{\ast}$&$SFR/M_{H_2}$\\
		                           &    Minutes    &   mK   &K  km s$^{-1}$&10$^7$K km s$^{-1}$ pc$^{-2}$&10$^8M_{\odot}$&&&$10^{-10}yr^{-1}$\\
		\hline
AGC 122966 &336&3.5& 0.23$\pm0.057$&2.2$\pm0.54$&$1.9\pm0.46$&$0.19\pm0.05$&$1.6\pm0.4$&1.3\\
                      &        &   &   $<$0.053$^c$  &$<$0.50 &$<$0.43 &$<$0.043&$<$0.36 &$>$5.7\\
%AGC 219533  &  63 &10.4& $<$0.16    &$<$1.8  & $<$1.5 &$<$0.094&$>$3.0\\
SECCO-dI-1&316 &5.2&$<$0.078 &$<$1.2 &$<$1.0&$<$0.083& $<$10.0&...\\
AGC 334315$^e$& 432&4.4&$<$0.066&$<$0.43&$<$0.37&$<$0.026 &$<$0.60&$>$12.1\\
\hline
UGC2162&107&9.6&$<$0.14&$<$0.027 &$<$0.023&$<$0.012&$<$0.12 &$>$37.8\\
SECCO-dI-2 &294 & 5.7&$<$0.086&$<$0.17 &$<$0.15&$<$0.071&$<$1.9 &... \\
Coma P$^e$&95 &19.6&$<$0.29 &$<$0.011&$<$0.0095&$<$0.027&$<$2.2&$>$3.3 \\
\hline
VLSB-A &121 &12.5&$<$0.19&$<$0.067&$<$0.058&...&&...&\\
		\hline
	\end{tabular}
	\end{center}
$a$. in T$_{mb}$  with frequency resolution of 0.195 MHz, which corresponds to $\sim$ 0.5 km s$^{-1}$ at 115GHz.  $b$. 3$\sigma$ upper limits for velocity integrated flux for 50 km s$^{-1}$ line width  in T$_{mb}$. $c$. 3$\sigma$ upper limit estimation for AGC 122966 if that feature is not real. $d$.   $\alpha_{CO}$=8.6$M_{\odot}$/K km s$^{-1}$pc$^{2}$ is used for estimating molecular gas mass.   $e$. The offsets of the pointing centers to the centers of  optical emission is about 15$''$ for  AGC 334315 and 9$''$ for Coma P,  which can cause under-estimation of the upper limits of CO emission in these two galaxies.
\end{table}

% Don't change these lines
\bsp	% typesetting comment
\label{lastpage}
\end{document}